\documentclass{article}

\usepackage[a4paper, margin=1in]{geometry}
\usepackage{authblk}
\usepackage{amsmath}
\usepackage{lineno}
\usepackage{graphicx}
\usepackage{epsfig}
\usepackage{float}
\usepackage{enumitem}

\usepackage[dvipsnames,svgnames,x11names]{xcolor}

\begin{document}
\title{Spin filtering on demand via localized states in an atomic-scale resonant tunneling magnetic tunnel junction}
\author[1,2]{M. Bazarnik}
\author[1,2]{A. Schlenhoff\thanks{schlenhoff@uni-muenster.de}}
\affil[1]{Institute of Physics, University of Münster, Münster, Germany}
\affil[2]{Department of Physics, University of Hamburg, Hamburg, Germany}
\maketitle
\begin{abstract}
Spin filtering and its back-action spin transfer torque (STT) are key ingredients of latest spintronic devices based on magnetic tunnel junctions (MTJs).
Resonant tunneling (RT), implemented by design or occurring as parasitic effects, is known to crucially affect macroscopic device performance, but direct experimental access to its individual microscopic processes has remained difficult. 
Here we apply the RT scheme from MTJs to spin-polarized scanning tunneling microscopy (SP-STM) for ultimate miniaturization obtained by addressing distinct sites on individual nanomagnets.
Combined with energy selectivity, our experimental model set-up enables to study the spin filtering capabilities of RT through an individual spin-split vacuum resonance state and of the corresponding STT exerted on the nanomagnet. 
We find, that the sign and magnitude of the STT follow the effective spin-polarization of the resonance state, which, as we show, can by tailored on demand either by adjusting the applied bias or the current injection point on the nanostructure. 
We anticipate, that our atomic-scale RT-MTJ approach and the discovery of a versatile tunable spin-filter at smallest scale will prove invaluable for studying and designing next generation MTJs potentially based on recently discovered 2D van-der-Waals magnets or altermagnets.
\end{abstract}

\begin{figure}[hbt!]
\begin{center}
\includegraphics[width=0.78\textwidth]{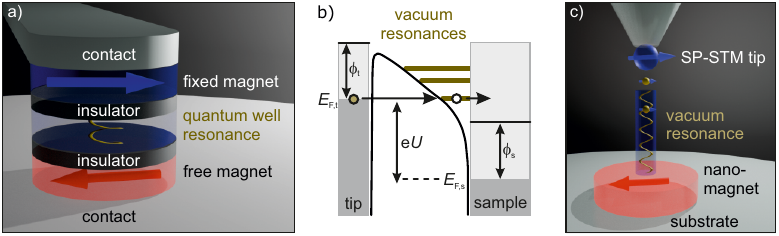}
\caption{\textbf{Resonant tunneling magnetic tunneling junction (RT-MTJ).}
\textbf{(a)}~Typical pentalayer design, composed of a central layer hosting quantum well resonances confined by adjacent insulating layers, sandwiched by ferromagnetic leads.
\textbf{(b)}~The resonant tunneling scheme in an STM set up. Upon selecting an appropriate bias voltage $U$, electrons from the probe tip tunnel resonantly through a vacuum resonance trapped in the potential well in front of the surface. ($\phi_{\rm t, s}$: tip and sample work function, $E_{\rm F,t, s}$: tip and sample Fermi level.)  
\textbf{(c)}~Atomic-scale analogue of an RT-MTJ. 
Due to the atomic-scale dimensions of the magnetic SP-STM tip apex, spatially localized spin-polarized electrons tunnel resonantly through a spin-dependent vacuum resonance into a nanomagnet on the substrate.  
}
\label{fig_RTMTJ}
\end{center}
\end{figure}
\section*{Introduction}
Spin filtering on demand is a key concept in complex spintronic devices, enabling the selective and dynamic manipulation of electron spins for advanced electronics and quantum technologies such as spin transistors, magnetic memories, and spin-based quantum computing devices.~\cite{Jia2025, Thapa2025, Shao2025, Hirohata_Review_2020}
By switching the filter on and off, or by choosing the polarity of the generated spin current, the electron's spin state encodes and processes quantum information.~\cite{Siwach_Quntum_Spin_2021, bordoloi_valve_2020}
Spin-polarized currents obtained by spin filters can be used to switch the magnetization direction of a soft, "free" magnet in a preferential direction by means of Spin Transfer Torque (STT), where spin angular momentum transfer from spin-polarized electrons exerts a "torque" on the soft magnet's magnetization.~\cite{berger_emission_1996, slonczewski_current-driven_1996, brataas_current-induced_2012}
On the one hand, STT based switching has been reported in atom-by-atom assembled quantum magnets~\cite{khajetoorians_current-driven_2013}, magnetic nano-skyrmions in thin-film systems~\cite{romming_writing_2013} and magnetic nanoislands~\cite{krause_current-induced_2007, herzog_heat_2010, schlenhoff_individual_2012} using spin-polarized scanning tunneling microscopy (SP-STM), demonstrating addressability at the atomic scale and thus the potential for ultimate miniaturization. 
However, in SP-STM the spin-polarization of currents and thus the STT is not well-controllable, depending on the characteristics of the individual probe tip.~\cite{Wortmann2001, wiesendanger_spin_2009}
A spin filter to be placed between the latter and the soft-magnetic surface to control the spin-polarization while maintaining the high locality of the technique is lacking so far.

On the other hand, STT is instrumental in layered magnetic tunneling junction (MTJ) devices used in latest commercial magnetoresistive random-access memory (STT-MRAM) cells.~\cite{worledge_spin-transfer_2024, igarashi_single-nanometer_2024, Hirohata_Review_2020}
Here, a high tunneling magnetoresistance (TMR) for reading and renitention of storing information, but a low switching current density for writing and thus a high STT efficiency is required.
These technological demands have prompted intensive efforts in complex stacking designs beyond the orginial trilayer (ferromagnet-insulator-ferromagnet) MTJ.~\cite{igarashi_single-nanometer_2024, worledge_snapshot_2023}  
The latter have gained particular momentum following the prediction of an enhancement of TMR and STT by means of resonant tunneling (RT) in pentalayer devices.~\cite{petukhov_resonant_2002, vedyayev_resonant_2006, theodonis_enhancing_2007, iovan_spin_2008, chatterji_enhancement_2014, chen_enhanced_2015, jia_enhanced_2016}
As illustrated in Fig.\ref{fig_RTMTJ}(a), such a RT-MTJ device is composed of a thin central (semiconducting, ferromagnetic or non-magnetic metallic) layer hosting quantum well resonances confined by adjacent insulating layers, that are themselves sandwiched by the ferromagnetic leads. 
Notably, also in standard trilayer MTJ resonant tunneling via discrete magnetic pinhole states~\cite{Victora_IEEE_2010} and impurity states~\cite{tsymbal_resonant_2003, birol_spin_2009, tang_effect_2011, harikumar_enhanced_2022}, respectively, is expected to induce a bias-dependent modification of the TMR and STT, that crucially depends on the state's energetic position.
While intensive theoretical work has been done on these resonant tunneling effects in (RT-)MTJ devices~\cite{lu_spin-dependent_2005, vedyayev_resonant_2006, theodonis_enhancing_2007, Victora_IEEE_2010, birol_spin_2009, mojumder_effect_2010, tang_effect_2011, chatterji_enhancement_2014, chen_enhanced_2015, harikumar_enhanced_2022, jiang_quantum-well_2025}, direct experimental access to the corresponding microscopic process itself has remained difficult.~\cite{tsymbal_resonant_2003, nozaki_quantum_2006, iovan_spin_2008, herranz_tunneling_2010}
This is mainly because the measurement of layered devices involves spatial averaging over an ensemble of discrete states, each of which exhibiting a slightly different energy due to the surface roughness of the interface between the center and the adjacent layers or to different chemical bonds in the case of impurity states.~\cite{Victora_IEEE_2010}

The concept of resonant tunneling via discrete states can be applied to an STM setup providing ultimate lateral resolution and the capability of selectively addressing an individual resonance state (RS) at a distinct energy.
As illustrated in Fig.~\ref{fig_RTMTJ}(b), discrete vacuum RSs evolve in the vacuum gap between the tip and sample, through which electrons tunnel resonantly when the tip's Fermi level is aligned with the RS's energy upon adjusting the applied bias voltage.~\cite{binnig_tunneling_1985, Gundlach_zur_1966} 
These RSs derive from image-potential states, which are unoccupied electronic states evolving in front of any polarizable surface, vertically confined by the Coulomb-like surface "image" potential and the reflection properties of the surface.~\cite{echenique_existence_1978, Hoefer2016}
Owing to the electric field between the tip and the sample, these states experience a Stark shift in the STM setup.~\cite{crampin_lifetimes_2005} 
Above magnetic surfaces, the states' binding energies depend on the orientation of their spin magnetic moment with respect to the underlying surface magnetization direction.~\cite{Donath_realization_2007, schlenhoff_vacuum_2019}
Since electrons in such a state move quasi-free parallel to the surface, its spin-splitting allows to create a spin-polarized, two-dimensional hot electron gas in front of the surface by selectively pumping one spin state.~\cite{Donath_realization_2007} 
Its spatial variation above complex surface spin textures has been recently studied by means of spin-polarized resonant tunneling in SP-STM.~\cite{schlenhoff_vacuum_2019, bazarnik_image-potential_2025}
Moreover, by means of the latter, reading and writing the magnetic state information of a nanomagnet at technically feasible nm-distances has been demonstrated.~\cite{schlenhoff_individual_2012, schlenhoff_real-space_2020}.

It is well known that nanostructures on surfaces can laterally localize RSs.~\cite{rejali_confined_2022, schouteden_quantum_2009, borisov_image_2007, fischer_image_1993}
Especially, at lateral boundaries of nanoislands, rim-localized RSs energetically shifted with respect to the nanoisland's undisturbed center RSs has been found.~\cite{stepanow_localization_2011}
While the emergence of these boundary-localized states has been established to be of general nature for any ad-island-substrate system with a sufficiently large variation of the work function between adsorbate and substrate,~\cite{stepanow_localization_2011} the question, how these RSs are affected by the magnetism of the underlying nanostructure has not been addressed so far.

In this work, we report on combining the resonant tunneling scheme from (RT-)MTJ devices with SP-STM, as illustrated in Fig.~\ref{fig_RTMTJ}(c), to study the spin filtering capabilities of resonant tunneling through an individual spin-split RS and of the corresponding STT exerted on the soft nanomagnet. 
We find that the STT follows the effective spin-polarisation of the RS hot-electron gas, including a bias-dependent sign inversion, all while maintaining the same polarity of bias, thus demonstrating the feasibility of bias-tunable spin-selective resonant tunneling in an atomic-scale RT-MTJ device.
Moreover, we show, how the energetic shift of RSs localized at lateral restrictions of nanomagnets affect the spatial variation of the effective hot-electron spin-polarisation and surprisingly find its spatial inversion at certain electron energies above the homogeneously magnetized nanostructure. 
In addition to the observed bias-tunability, this opens the path to spin filtering on demand by tuning the lateral position of current injection on nanostructured surfaces. 

%
\section*{Results}
\section*{Spin-polarization of boundary-localized vacuum resonances}
The spatial delimitation of a nanostructure causes a disturbance of its RS resulting in the evolution of a rim-localized state energetically shifted with respect to the undisturbed center state of the respective nanostructure.~\cite{stepanow_localization_2011}
The question arises, how these vacuum RSs - center and rim state - are affected by the magnetism of the nanostructure.
To answer this, we study two well-known magnetic systems: arbitrarily-shaped in-plane magnetized Fe islands on W(110) and triangular out-of-plane magnetized Co islands grown on Ir(111).
%
\begin{figure}[hbt!]
\begin{center}
\includegraphics[width=0.5\columnwidth]{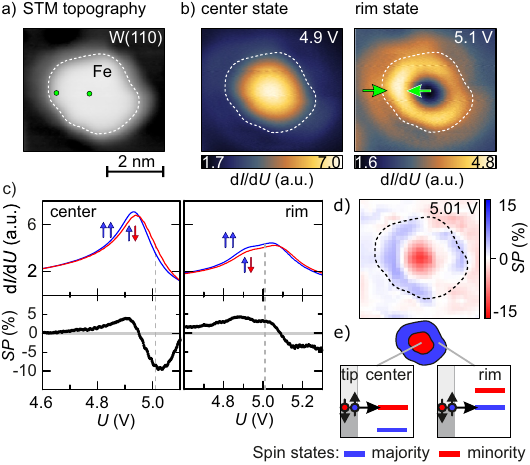}
\caption{\textbf{Spin-polarization of a nanomagnet's center and rim vacuum resonance state.}
\textbf{(a)}~STM topography of an ML Fe/W(110) nanomagnet ($U=0.1$\,V). 
\textbf{(b)}~d$I$/d$U$ maps recoded at $U=4.9$\,V and $5.1$\,V, respectively, revealing the center state (left) and the rim state (right). 
Spatial extent of the nanoisland as determined from (a) is indicated. 
The bright region (marked by arrows) in the right map reveals the spatial extent of the rim state.
\textbf{(c)}~Top: Spin-resolved d$I$/d$U(U)$ curves, taken on the center (left) and the rim (right) as indicated in (a), in parallel $\uparrow\uparrow$ (blue) and antiparallel $\uparrow\downarrow$ (red) alignment of tip and island magnetization, respectively.
Bottom: Spin-polarization $SP(U)$ of the hot-electron gas on the center (left) and the rim (right), respectively. 
\textbf{(d)}~$SP$ map at $U=5.01$\,V (bias voltage marked by dashed lines in (c)).
\textbf{(e)}~Top: Schematic of the nanomagnet with rim and center $SP$, as observable in (d). 
Bottom: Corresponding schematic of the spin-dependent tunneling at the island center and rim, respectively. 
($I=2$\,nA, $f=3980$\,Hz, $U_{\rm {Pk-Pk}}=120$\,mV, $T=36$\,K.)
}
\label{fig_SP_Island}
\end{center}
\end{figure}
%
An STM topography of a typical monolayer (ML) Fe/W(110) nanomagnet is shown in Fig.~\ref{fig_SP_Island}(a). 
d$I$/d$U$ maps recoded at $U=4.9$\,V and $5.1$\,V corresponding to resonant tunneling via the center and the rim state, respectively, are shown in Fig.~\ref{fig_SP_Island}(b).
Clearly observable is the highest probability density for the center state right in the nanoisland center, and a ring-shaped probability density for the rim state, analogue to previous findings for Li clusters on Cu(100)~\cite{stepanow_localization_2011}.  
From the d$I$/d$U$ map at $U=5.1$\,V, the lateral extension of the rim state is estimated to be about $1$\,nm.
ML Fe/W(110) nanomagnets are known to be monodomain particles with an uniaxial anisotropy lying in the surface plane~\cite{elmers_FeML_1990, krause_current-induced_2007}.  
In Fig.~\ref{fig_SP_Island}(c), spin-resolved constant current d$I$/d$U(U)$ curves on the center and the rim of the nanomagnet, respectively, with parallel and antiparallel alignment of tip and island magnetization are shown. 
For the spectroscopy curves recorded on the center, a pronounced maximum at $U=4.9$\,V is observed, indicating the energetic position of the center state. 
For the curves recorded at the rim, a double-peak like feature is found. 
Here, the global maximum at $U=5.08$\,V indicates the rim state energy, while the lower-energy shoulder stems from some remaining tunneling into the center state. 
The boundary-induced rim state is found at an higher energy than the RS on the island center (see supplementary note 1 for more details). 
Note, that we observe a similar situation at the rim of extended ML Fe/W(110) nano-stripes, demonstrated in supplementary note 1.

Interestingly, for both states the peak in the spin-dependent curves changes position and height for the two magnetic configurations of the underlying nanomagnet, indicating a spin-polarization of both the center and the rim state~\cite{schlenhoff_individual_2012}. 
Thus, not only the center RS but also the rim state exhibits a significant exchange splitting, as schematically depicted in Fig.~\ref{fig_SP_Island}(e). 
A spin-polarization of (center) vacuum RSs on this and other types of magnetic nanostructures has been observed before~\cite{schlenhoff_individual_2012, kubetzka_spin-polarized_2007, bazarnik_image-potential_2025}.
However, spin-resolved studies have yet ignored boundary effects.  
Moreover, beyond the finding of a spin-polarized boundary-localized resonance, an interesting yet unexplored phenomenon evolves, which is based on the energetic shift between the spin-split rim and center state. 
This is best observable upon determination of the effective bias-dependent spin-polarization of the hot-electron gas above the nanomagnet, given by the quantity $SP(U)=($d$I/$d$U_{\uparrow\uparrow}(U)-$d$I/$d$U_{\uparrow\downarrow}(U))/($d$I/$d$U_{\uparrow\uparrow}(U)+$d$I/$d$U_{\uparrow\downarrow}(U))$~\cite{bazarnik_image-potential_2025}.  
It is shown as a function of sample bias in Fig.~\ref{fig_SP_Island}(c) for the rim state as well as for the center state. 
For both curves, $SP(U)$ changes sign, when the respective d$I/$d$U_{\uparrow\uparrow}(U)$ and d$I/$d$U_{\uparrow\downarrow}(U)$ curves cross each other. 
Note, that this is analogue to the expected bias-dependent sign change of the TMR for resonant tunneling through an individual state in (RT-)MTJ devices.\cite{Liu2013, iovan_spin_2008,Victora_IEEE_2010}
In the case examined here, majority (spin-up) and minority (spin-down) RSs are expected to intrinsically overlap, being both accessible for a fixed electron energy e$U$ and contributing both to the tunneling process~\cite{bazarnik_image-potential_2025, schlenhoff_individual_2012}.
However, the weight of the respective contributions is bias-dependent (see detailed discussion in Ref.~\cite{bazarnik_image-potential_2025}). 
A crossing point in the spin-dependent spectroscopy curves marks an energy e$U_{\rm cross}$ at which the tunneling process changes, e.g. from majority dominated (at lower energies) to minority dominated (at higher energies), and thus a sign reversal in the effective spin-polarization of the hot-electron gas above the nanomagnet.
Since the spin-split center and the rim state peak are shifted in energy with respect to each other, the crossing point for the center state is at $U=4.94$\,V, but for the rim state at $5.08$\,V. 
As a result, energies are found, where the two states exhibit opposite spin-polarization.  
For example, at a bias voltage of $U=5.01$\,V, $SP<0$ for the center, but $SP>0$ for the rim state. 
Being located at different regions on the nanomagnet, this implies a laterally varying spin-polarization of the hot RS electron gas above the nanomagnet.
Indeed, the $SP$~map at $U=5.01$\,V shown in Fig.~\ref{fig_SP_Island}(d), reveals a negative effective spin-polarization in the center of the nanomagent, while the rim is positively polarized, as schematically sketched in Fig.~\ref{fig_SP_Island}(e).
Remarkably, we find this spatial variation of the effective spin-polarization of the hot-electron gas, although the underlying nanomagent exhibits a uniform magnetization. 
Fe/W(110) nanomagnets are well-studied systems and neither spin-polarized edge states at lower energies, nor structural or electronic changes at the nanomagnet's rim or other phenomena, which might result in similar observations, are known.  
Hence, the spatial variation of the effective spin-polarization of the hot-electron gas above the homogeneously magnetizated nanostructure is solely originating from the evolution of the boundary-localized vacuum resonance and its spin-polarization. 
%
\begin{figure}[htb!]
	\centering
		\includegraphics[width=0.75\textwidth]{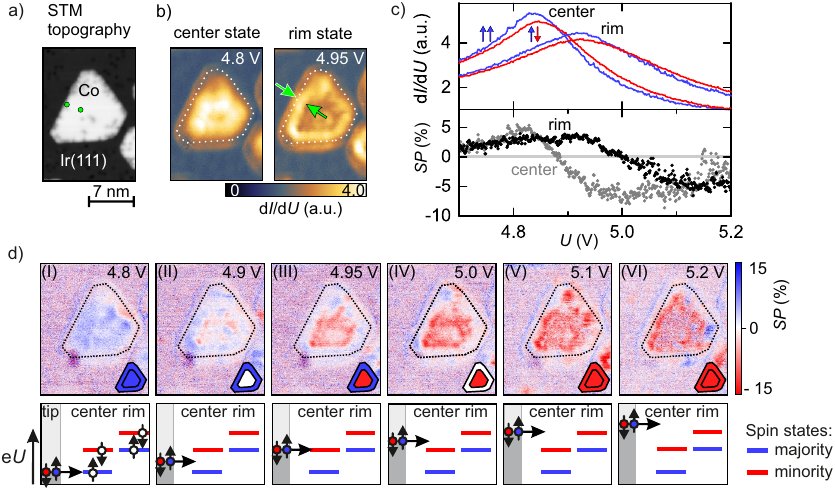}
		\caption{\textbf{Energy dependence and spatial modulation of a nanomagnet's hot-electron RS spin-polarization.}
\textbf{(a)}~STM topography of a ML Co/Ir(111) nanomagnet ($U=-0.5$\,V). 
\textbf{(b)}~d$I$/d$U$ maps recoded at $U=4.8$\,V and $4.95$\,V, respectively, revealing the center state (left) and the rim state (right). 
Spatial extent of the nanoisland as determined from (a) is indicated. 
\textbf{(c)}~Top: Spin-resolved d$I$/d$U(U)$ curves, taken on the center and the rim as indicated in (a), in parallel $\uparrow\uparrow$ (blue) and antiparallel $\uparrow\downarrow$ (red) alignment of tip and island magnetization, respectively. 
Bottom: Spin-polarization $SP(U)$ of the hot-electron gas on the center and the rim, respectively. 
\textbf{(d)}~Top: Bias-dependent series of SP maps (with schematic of the rim and center SP in the inset). Bottom: Schematic of the spin-dependent tunneling as a function of electron energy e$U$. 
($I=0.1$\,nA, $f=3777$\,Hz, $U_{\rm {Pk-Pk}}=10$\,mV, $T=6.5$\,K, $B=\pm1$\,T.)
}
\label{fig_Co}
\end{figure}
\section*{Spatial and energy dependence of the resonances' spin-polarization}
In order to investigate the spatial and energy dependence of the effective spin-polarization of the hot-electron gas above monodomain nanomagnets in more detail, the system of ML Co nanoislands on Ir(111) was chosen. 
Since the latter grow in well-defined geometrical shapes, the formation of the boundary-induced rim state is expected to be even more pronounced than on the arbitrarily shaped Fe/W(110) nanomagnets. 
The two-dimensional nanoscale Co/Ir(111) islands are ferromagnetic and single domain, with the magnetic easy axis normal to the surface~\cite{bickel_magnetic_2011}. 
An STM topography of a typical ML Co/Ir(111) nanomagnet is shown in Fig.~\ref{fig_Co}(a). 
In Fig.~\ref{fig_Co}(b), two d$I$/d$U$ maps recorded at $U=4.8$\,V and $4.95$\,V, respectively, are shown, revealing the center and the rim state, respectively, on the Co nanomagnet. 
From the d$I$/d$U$ map at $U=4.95$\,V, the lateral extension of the rim state is estimated to be about $2$\,nm. 
A spin-resolved d$I$/d$U(U)$ spectroscopy on the center and the rim of the island, respectively, with parallel and anti-parallel alignment of tip and island magnetization is shown in Fig.~\ref{fig_Co}(c). 
First, it reveals the boundary-induced rim state (maximum at $U=4.92$\,V), being higher in energy than the RS on the island center (maximum at $U=4.84$\,V), similar to the ML Fe/W(110) system. 
Second, for both states the peak changes position and height, depending on the relative magnetic orientation of the tip and the nanoisland. 
Consequently, also on this type of nanomagnet, not only the center RS but also the rim state exhibits a significant exchange splitting. 
The resulting effective spin-polarization $SP(U)$ of the hot RS electron gas is shown in Fig.~\ref{fig_Co}(c) for both the rim state and the center state. 
Analogue to our observation on Fe/W(110) islands, for both curves the spin-polarization changes sign, when the respective d$I/$d$U_{\uparrow\uparrow}(U)$ and d$I/$d$U_{\uparrow\downarrow}(U)$ curves cross each other, indicating the energy e$U_{\rm cross}$ at which the tunneling process changes from majority dominated at lower energies to minority dominated at higher energies. 
Since the center and the rim state peak are shifted in energy with respect to each other, the crossing point for the center state is at $U=4.88$\,V, but for the rim state at $5$\,V. 
This results in an energy range of about $120$\,meV, where the two states, localized at different regions on the nanomagnet, exhibit opposite spin-polarization.

Bias-dependent $SP$ maps were computed from d$I$/d$U$ maps taken for parallel (d$I/$d$U_{\uparrow\uparrow}$) and antiparallel (d$I/$d$U_{\uparrow\downarrow}(U)$) alignment of tip and island magnetization, respectively. 
They reveal the spatial distribution of the effective spin-polarization of the hot-electron gas above the nanomagnet as a function of electron energy (see Fig.~\ref{fig_Co}(d)). 
As long as the electron energy is lower than for both the center and the rim state, the spin-polarized tunneling from the tip into the spin-split RSs is dominated all over the island by the majority RSs (see schematic in the bottom of Fig.~\ref{fig_Co}(d)~(I)). 
Consequently, at these electron energies the hot-electron gas exhibits a positive spin-polarization above the entire nanomagnet, resulting in a blue color in the map. 
When going higher in energy~(II), spin-polarized electrons tunnel equally in the majority and the minority spin state at the center, resulting in a vanishing spin-polarization (white center in the map), whereas at the rim the tunneling is still majority-dominated (blue in Fig.~\ref{fig_Co}(d)). 
When increasing further the electron energy~(III), the spin-polarization of the center RS changes sign, indicated by the red color in the map, whereas $SP$ of the rim state is still positive~(blue). 
Consequently, we find an opposite spin-polarization of the hot-electron gas above the rim and the center of the nanomagnet. 
Going higher in energy ~(IV), also the crossing point of the rim state is met, where tunneling occurs equally in its majority and minority spin state, resulting in a vanishing spin-polarization (white rim in the map). 
On further increasing the energy~(V), the red center region of the maps is growing towards the island edges, since also the rim state gets more and more minority-spin polarized with increasing electron energy. 
Finally, when the electron energy~(VI) is higher than for both the center and the rim state, the spin-polarized tunneling is dominated all over the island by the minority RSs and the hot-electron gas exhibits a negative spin-polarization above the entire nanomagnet.\\
\indent In conclusion, the spatial limitation of ferromagnetic, monodomain nanoislands causes independent on their shape and easy axis orientation, respectively, the evolution of an exchange-split RS that is bound to the rim, leading to a spatially varying spin-polarization of the hot-electron gas above uniformly magnetized nanoislands. 
The spatial modulation of this spin-polarization is strongly bias-dependent, resulting even in a sign change for a certain range of electron energies.
In this range, as a rule of thumb, tunneling occurs at the falling flank of the center peak, but at the rising flank of the rim peak, as exemplarily marked in Fig.~\ref{fig_SP_Island}(c). 
Consequently, this lateral inversion of the spin-polarization is expected to exist on any nanomagnet, that exhibits a sufficient large energetic shifting of the rim state with respect to the center RS, resulting in a tunneling process that is majority dominated at the rim, but minority dominated at the center. 
\section*{Vacuum resonances as tunable spin-filter for spin-current induced magnetization switching}
\begin{figure}[htb!]
\begin{center}
\includegraphics[width=0.5\columnwidth]{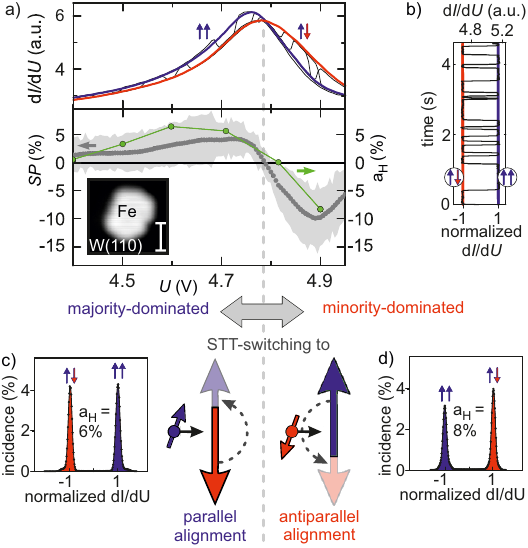}
\caption{\textbf{Tailoring the spin-transfer torque (STT) by tunneling through spin-polarized vacuum resonances.}
\textbf{(a)}~Inset: Constant current image of a thermally switching Fe/W(110) nanomagnet ($U=-0.2$\,V). Scale bar is $2$\,nm. 
Top: Spin-resolved d$I$/d$U(U)$ curves, taken on the nanomagnet's center, revealing the spin-split RS. 
The signal jumps between the curves for parallel $\uparrow\uparrow$ (blue) and antiparallel $\uparrow\downarrow$ (red) alignment of tip and island magnetization, due to frequent switching events. 
Bottom: Corresponding spin-polarization $SP(U)$ of the hot-electron gas on the center (dark grey) with measurement uncertainty (light grey). 
The crossing point of d$I$/d$U(U)_{\uparrow\uparrow}$ and d$I$/d$U(U)_{\uparrow\downarrow}$ and the respective inversion of SP is marked by the dashed line.
Current-induced asymmetry~$a_{\rm H}$ (green) between the two magnetization states of the island as a function of bias~$U$.  
\textbf{(b)}~Telegraph noise, recorded on the nanomagnet's center within the first RS ($U=4.7$\,V). 
\textbf{(c)}~State histogram, as determined from the telegraphic noise recorded at $4.7$\,V and \textbf{(d)}~$U=4.9$\,V, respectively. 
Schematic sketch: Upon sign inversion of $SP(U)$, the STT reverses accordingly, favoring a parallel (antiparallel) alignment of tip and sample magnetization for positive (negative) SP, as deduced from $a_{\rm H}$.
($I=2$\,nA, $f=4333$\,Hz, $U_{\rm {Pk-Pk}}=120$\,mV, $T=39,5$\,K.)
}
\label{figSTT}
\end{center}
\end{figure}
When the hot electron-spins tunnel through the RSs into the surface, a STT is exerted on the sample~\cite{schlenhoff_individual_2012}.  
It has been demonstrated that, by observing their switching behavior, ML Fe/W(110) nanomagnets serve well as STT analyzers~\cite{schlenhoff_individual_2012, krause_joule_2011}. 
Hence, this sample system is well suited for a detailed investigation of the influence of the RSs' spin polarization on the current-induced magnetization switching and for testing the RSs' aptitude to serve as spin filter.

In Fig.~\ref{figSTT}(a) an SP-STM constant current map of a typical ML Fe/W(110) nanomagnet is shown. 
A faint stripe pattern is visible on the nanomagnet, reflecting the thermally induced switching of its magnetization when imaged line by line. 
Positioning the tip stationary above the nanomagnet's center and recording the magnetic d$I$/d$U$ signal as a function of time allows for the observation of its temporal magnetization evolution.  
In Fig.~\ref{figSTT}(b), a section of the d$I$/d$U(t)$ raw signal is shown, recorded by resonant tunneling through the first RS. 
A characteristic telegraph noise is observed, with the two levels indicating the parallel and antiparallel alignment of the nanomagnet’s magnetization with respect to the stable tip magnetization~\cite{schlenhoff_individual_2012, krause_current-induced_2007}.  
The RS induced peak in the spectroscopy is shown in Fig.~\ref{figSTT}(a). 
Several d$I$/d$U(U)$ curves are shown, with each of them exhibiting sudden jumps. 
Plotting them on top of each other reveals that the jumps indicate switching events of the nanomagnet, with the RS peak slightly changing in terms of position and height. 
As discussed above, this finding indicates intrinsically overlapping spin-split RSs that reflect the magnetization of the underlying nanomagnet~\cite{schlenhoff_individual_2012}. 
The resulting effective spin-polarization $SP$ as a function of sample bias~$U$ is shown in the bottom of Fig.~\ref{figSTT}(a).
It changes sign at e$U_{\rm cross}$ (marked by the dashed line in Fig.~\ref{figSTT}(a)), at which the tunneling process changes from majority dominated at lower energies to minority dominated at higher energies.

The question arises, whether this bias-dependent change of the effective spin-polarization of the resonant tunneling electrons affects the switching behavior of the underlying magnet.
In order to study this, the telegraph noise d$I$/d$U(t)$ has been recorded on the nanomagnet for different bias voltages from $U=4.4$\,V to $4.9$\,V, spanning the energy range of resonant tunneling into the first spin-split RS and covering the transition from positive to negative $SP(U)$, e.g. from majority-dominated to minority-dominated tunneling. 
From the telegraph noise data, a histogram of the two states of the nanomagnet for each sample bias $U$ has been determined. 
Figure~\ref{figSTT}(c) and (d) show exemplary histograms for $U=4.7$\,V (with $SP>0$) and $U=4.9$\,V (with $SP<0$), respectively. 
For both cases a histogram asymmetry~$a_{\rm H}$ is observed. 
However, with $SP>0$ the parallel alignment of the nanomagnet’s magnetization with respect to the tip magnetization is favored, whereas with $SP<0$ the antiparallel alignment is favored. 
The favoring of one magnetic orientation of the nanomagnet at the cost of the other is a clear indication of a current-induced STT~\cite{schlenhoff_individual_2012, krause_joule_2011}. 
Clearly, the STT changes its sign, when the polarity of $SP$ changes. 
Consequently, upon selecting a bias on the rising or falling flank of the spin-split RS peak, respectively, the magnetization can be switched back and forth on demand.
\
In Figure~\ref{figSTT}(a) the current-induced state asymmetry~$a_{\rm H}(U)$ as deduced from all the bias-dependent telegraph noise data, is shown. 
Its absolute value and polarity is a measure for the magnitude and the direction of action of the current-induced STT. 
Evidently, $a_{\rm H}(U)$ follows the evolution of $SP(U)$.
This indicates that the vacuum RSs in front of the surface serve as a spin-filter that can be used to tailor the effective spin-polarization of the hot-electron gas by adjusting the applied bias, and thus the STT for current-induced magnetization switching.

Interestingly, our experiments indicate, that the sign reversal in $SP(U)$ and STT is not accompanied by a corresponding sign reversal in the spin-polarization of the total current (see SOM note 2). 
These findings are in line with theoretical studies in that the disparate behavior of the spin-polarized currents and the STT has been identified as a clear fingerprint of spin-dependent resonant tunneling through spin-split discrete states with majority and minority states being close in energy,~\cite{theodonis_enhancing_2007} as is the case in our experiments.
Going beyond these theoretical predictions, we further experimentally demonstrate that it is the effective hot-electron spin-polarization $SP(U)$ that drives the resonant tunneling mediated STT.

Moreover, based on the finding of a spatial inversion of $SP(U)$ on the nanoisland's rim with respect to the center for certain bias voltages, as demonstrated in Fig.~\ref{fig_SP_Island} and Fig.~\ref{fig_Co}, the STT is expected to reverse correspondingly at the rim with respect to the center. 
Consequently, by means of resonant tunneling through RSs in front of the surface, the STT for current induced magnetization switching can by tailored on demand either by adjusting the applied bias or the current injection point on the nanostructure, with the RS serving as viable spin filter.
\section*{Discussion}
These proof-of-principle experiments expand the portfolio of current-induced magnetization switching on a very local scale by the discovery of a viable tunable spin filter located in the vacuum gap between the SP-STM probe tip and the magnetic sample. 
Our experimental approach further provides access to the microscopic processes of spin-dependent resonant tunneling inherent in (RT-)MTJ devices, linking a bridge between theoretical calculations and macroscopic observations, which is essential for advancing spin-dependent quantum resonant tunneling applications. 
We demonstrate the spin-selectivity of the resonant tunneling process, thereby opening the path to bias-selectivity of the STT, and interestingly find the effective hot-electron spin-polarization $SP(U)$ to be the driving factor.
Our experiments show-case how a lateral shift in the RS's energy affect the $SP(U)$ and thus the magnitude and direction of action of the STT, over that a typical (RT-)MTJ device would spatially average, potentially resulting only in an overall decrease of the macroscopic observables.
We have tested the mechanism on both in-plane and out-of-plane magnetized systems proving its general validity and versatility.
From a fundamental point of view, our finding of a spin-polarized RS localized at the spatial boundary of nanostructures, surprisingly causing a spatial inversion of $SP(U)$ above the uniformly magnetized nanostructure, provides novel insights in the physics of this special class of unoccupied surface states. 
Since the existence of the rim state is of general nature and, as we showed, independent from the nanostructure's shape and symmetry, a spatial variation of $SP(U)$ is expected at the boundary of any ferromagnetic adlayer-substrate system with a sufficiently large variation of the work function between adsorbate and substrate.
Along with our finding of a correlated inversion of the STT with $SP(U)$ this opens the path to tailor the local STT upon nanostructuring a magnetic film on a surface. 
Moreover, we expect that the modulation of the STT upon local resonant current injection can be utilized also on magnetic 2D van-der-Waals hybrid- and heterostructures, for that we have recently found a moiré-induced spatial variation of the hot-electron spin-polarization $SP(U)$.~\cite{bazarnik_image-potential_2025}
We believe that the microscopic access provided by our atomic-scale RT-MTJ approach will prove invaluable in future explorations of the potential of recently discovered 2D van-der-Waals magnets~\cite{Jia2025, Thapa2025, Wang2022, Mak2019} and altermagnets~\cite{Smejkal2022} for next generation MTJ devices.

\section*{Methods}
Experiments were performed in two home-build ultra high vacuum systems equipped with either a variable temperature scanning tunneling microscope (VT-STM) operated at temperatures close to $T=36$\,K or a cryogenic STM operated at a temperature of $T=6.5$\,K with a magnetic field normal to the sample surface of up to $B=6.25$\,T. 
W(110) was prepared by repeated cycles of glowing in oxygen atmosphere followed by a flash annealing at temperatures above $T>2400$\,K.  
0.06 ML of Fe were deposited at room temperature leading to ML nanoislands of different sizes on the W terraces. 
ML Fe/W(110) nanomagnets are known to be monodomain particles with an uniaxial anisotropy lying in the surface plane~\cite{elmers_FeML_1990, krause_current-induced_2007}
At a temperature around $T=36$\,K, these islands are superparamagnetic.
Therefore they invert their magnetic orientation from time to time due to thermal agitation~\cite{krause_magnetization_2009}. 
For recording the spin-resolved spectroscopy presented in Fig.~\ref{fig_SP_Island} the probe tip has been positioned on the island and a spectroscopy grid has been measured for both magnetic configurations before and after a switching event of the slowly switching nanomagnet.
However, in the case of a rapidly switching island, such as the one presented in Fig.~\ref{figSTT}, the probe tip has been positioned on the island and several consecutive spectra have been recorded.
During each spectroscopy, the island switched several times and therefore each individual spectra (black in Fig.~\ref{figSTT}(a)) is a mixture of two traces. \cite{schlenhoff_individual_2012} 
Therefore, after acquisition the data is binned for each bias value, separating the points into two spectra for parallel and anti-parallel configurations of the tip-sample magnetization. 
The binned data is then averaged (resulting in the blue and red curve in Fig.~\ref{figSTT}(a)), and spin asymmetry~$SP(U)$\cite{bazarnik_image-potential_2025} can be computed.

Ir(111) was cleaned by repeated cycles of Ar$^{+}$ sputtering ($800$\,V, $5$E$-6$\,mbar), annealing at temperatures ranging from $T=900$\,K to $1500$\,K in the presence of oxygen and a flash annealing at $T\approx 1500$\,K.
Subsequently $0.2$\,ML has been deposited onto the substrate held at room temperature.
The two-dimensional nanoscale Co/Ir(111) islands are ferromagnetic and single domain, with the magnetic easy axis normal to the surface~\cite{bickel_magnetic_2011}.
At the measurement temperature of $T=6.5$\,K, the Co nanomagnets are thermally stable. 
Remarkably, the coercitivity of these islands is greater than $4$\,T and magnetic saturation of the islands requires an applied field of at least $5$\,T. 
Using a thin-film ferromagnetic Fe/W-tip, the tip magnetization direction is controlled by applying a magnetic field of $B=\pm 1$\,T, which is sufficient to mostly align the tip magnetization in the out-of-plane direction.
When the applied field polarity is inverted, the alignment of the magnetic tip is inverted, allowing for investigating these Co islands in their parallel and antiparallel magnetic alignment with respect to the out-of-plane component of the tip magnetization, as has been already demonstrated by SP-STM~\cite{bickel_magnetic_2011}.
Hence, we used a soft ferromagnetic Fe coated W-tip for data acquisition presented in Fig.~\ref{fig_Co}, while antiferromagnetic Cr tips, sensitive to the in-plane component of the sample magnetization, were used for the data presented in Fig.~\ref{fig_SP_Island} and Fig.~\ref{figSTT}.
A lock-in detection technique was used to obtain d$I$/d$U$ maps and point spectroscopy data adding a small AC modulation to the bias voltage~$U$, the respective values are given in figure captions.
The d$I$/d$U$ maps were recorded simultaneously with the STM topography in the constant-current mode.
High bias differential conductance spectra have been recorded while maintaining a constant current setpoint during the bias voltage sweep~\cite{binnig_tunneling_1985, becker_electron_1985, schlenhoff_real-space_2020}. 

\section*{Acknowledgments}
The authors would like to thank R.~Wiesendanger for support and provision of laboratories and infrastructure. 
AS acknowledges financial support from the Deutsche Forschungsgemeinschaft via Grant Nos. SCHL2096/1-2 and SCHL2096/1-3.

\section*{Author Contributions}
AS conceived the experiments and lead the project.
Both authors performed the measurements and analysis and wrote the manuscript.

\bibliographystyle{naturemag}
\bibliography{STTviaIPS_main}

\end{document}


\maketitle

\section*{Note 1: Evolution of rim-localized vacuum resonances via nanostructuring}
\begin{figure}[hbt]
\begin{center}
\includegraphics[width=0.99\textwidth]{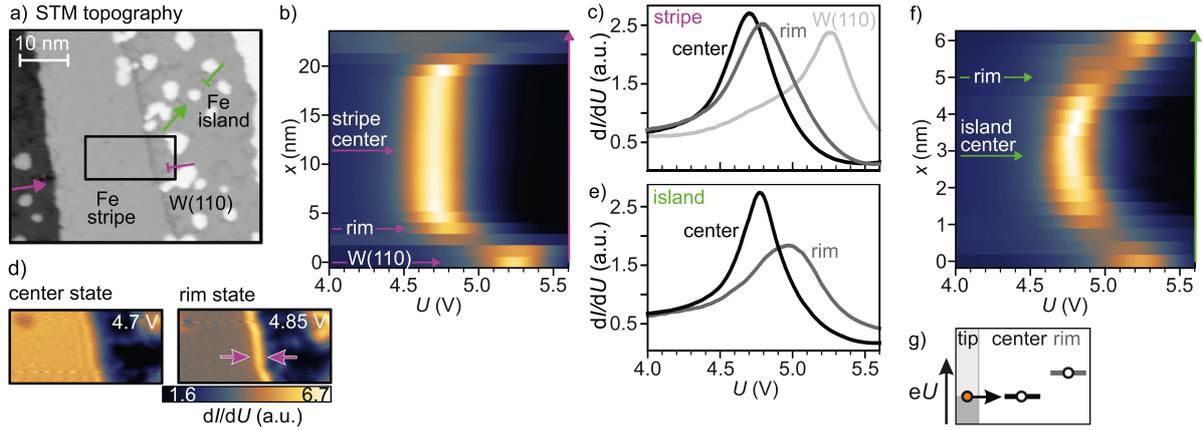}
\caption{\textbf{Evolution of rim-localized vacuum resonances via nanostructuring.}
\textbf{(a)}~STM topography image of ML Fe/W(110) stripes and islands ($U=0.1$\,V).
\textbf{(b)}~Constant current d$I$/d$U(U, x)$ recorded along the line marked in (a) by the purple arrow, revealing the spatial evolution of the first RS between the W(110) substrate and the ML Fe stripe. 
\textbf{(c)}~Exemplary d$I$/d$U(U)$ curves taken on the stripe center, at the rim and the substrate, respectively, as indicated in (b).
\textbf{(d)}~d$I$/d$U$ maps recorded at bias voltages corresponding to the respective RS peak position on the stripe center ($U=\,4.7$\,V, left) and rim ($U=\,4.85$\,V, right), as deduced from (c). 
Location of the maps is marked in (a) by the black frame. 
The bright region (marked by arrows) in the right map reveals the spatial extent of the rim state.
\textbf{(f)}~Constant current d$I$/d$U(U, x)$ recorded along the line marked in (a) by the green arrow, revealing the spatial evolution of the first RS between the substrate and the ML Fe island. 
\textbf{(e)}~Exemplary d$I$/d$U(U)$ curves taken on the island center and the rim, respectively, as indicated in (f). 
\textbf{(g)}~Schematic sketch: The rim-localized state of the nanostructures is found at higher energies than the center state, as deduced from the spectroscopy data shown in b)-f).
($I=2$\,nA, $T=36$\,K.)
}
\label{figRimState}
\end{center}
\end{figure}

In Figure~\ref{figRimState}(a) an STM topography image of a W(110) surface decorated with ML Fe stripes and islands is shown. 
d$I$/d$U(U)$ spectroscopy curves taken along a $23$\,nm long line across the Fe/W(110) stripe, with start and ending points on the lower and upper W(110) terrace, respectively, are shown in Figure~\ref{figRimState}(b).
The laterally resolved spectroscopy reveals the spatial evolution of the first vacuum resonance peak (manifested as bright feature in the waterfall plot) between the W(110) substrate and the ML Fe stripe.
In Fig.~\ref{figRimState}(c), three exemplary d$I$/d$U(U)$ spectroscopy curves from (b), recorded on the stripe center, at the rim of the stripe and on the substrate, respectively, are shown.
The maximum in the d$I$/d$U(U)$ curves reveals the energetic position of the respective resonance state (RS). 
In the center of the extended Fe ML stripe, the state is observed at lowest energy ($U=4.7$\,V), whereas on the W substrate it is found at highest energy ($U=5.28$\,V), which reflects the work function difference between the Fe ML and the substrate.
Evidently, at the rim of the Fe stripe, the state is found at an energy higher than for the undisturbed state at the film center.
Consequently, the spatial delimitation of the Fe film causes a disturbance of its RS resulting in the evolution of a state at the rim of the film, that will be called in the following 'rim state'.

From the spectroscopy curves shown in Fig.~\ref{figRimState}(c), at a fixed sample bias $U$ a d$I$/d$U(U)$ value that changes with lateral position is expected when scanning over the Fe stripe to the substrate, allowing to map the spatial extent of the states. 
Figure~\ref{figRimState}(d) shows two d$I$/d$U$ maps with their corresponding topography covering the transition region from the Fe stripe center to the adjacent W(110) terrace, as marked in Fig.~\ref{figRimState}(a). 
In the map recorded at $U=4.7$\,V, the stripe appears bright but a narrow region close to the rim appears slightly darker.
Here, the presence of the rim state results in a lower d$I$/d$U$ signal, since the chosen voltage is off-resonant with respect to the rim state peak position (compare Fig.~\ref{figRimState}(c)). 
In contrast, in the d$I$/d$U$ map at $U=4.85$\,V the rim state is imaged in resonance and the center state is off-resonant, resulting in a small region on the stripe at the rim appearing much brighter than the rest of the stripe area. 
From the map, we deduce the spatial probability density of the rim state to be localized within approx. $2$\,nm at the rim. 

Figure~\ref{figRimState}(f) shows d$I$/d$U(U)$ spectroscopy curves taken along a $6.1$\,nm long line across a ML Fe/W(110) island, with start and ending points on the W(110) substrate, as marked in the topography in Fig.~\ref{figRimState}(a).
Exemplary d$I$/d$U(U)$ spectroscopy curves from (f), recorded on the island center and on the island rim, respectively, are shown in Fig.~\ref{figRimState}(e). 
Again, the RS on the Fe nanostructure center is found at lower energies then on the W substrate (compare Fig.~\ref{figRimState}(c)). 
However, the center state energy is slightly higher $U=4.77$\,V on the Fe island compared to the extended Fe stripe. 
This energetic shift results from the lateral quantum confinement of the RS by the two-dimensional potential well on the surface formed by the nanoisland with its lower workfunction compared to the surrounding extended substrate.
As a consequence, the center state energy scales with the island size, and approaches only in the limit of very large islands the value for the extended stripe (not shown here).
While this has not been studied for Fe/W(110) nanoislands before, it is a well-studied phenomenon for various other material systems.~\cite{rejali_confined_2022, schouteden_quantum_2009, borisov_image_2007, fischer1993}
In analogy to the observation for the rim of the extendend stripe, also on the island rim a RS is found at an energy larger then on the nanostructure's center, as schematically sketched in Fig.~\ref{figRimState}(g).

In conclusion, the spatial limitation of a thin film, caused by its nanoscale restriction in one or two dimensions on a substrate, causes a disturbance of its RS resulting in the evolution of a boundary-localized state energetically shifted with respect to the undisturbed center state of the respective nanostructure, as demonstrated here for the rim of extended ML Fe/W(110) stripes and of ML Fe/W(110) nanoislands. 
As discussed in Ref.~\cite{stepanow_localization_2011}, the emergence of these states is of general nature for any adlayer-substrate system with a sufficiently large variation of the work function between adsorbate and substrate.

\section*{Note 2: Bias-dependence of the spin-polarized resonant tunneling current}
\begin{figure}[!hbt]
\begin{center}
\includegraphics[width=0.5\columnwidth]{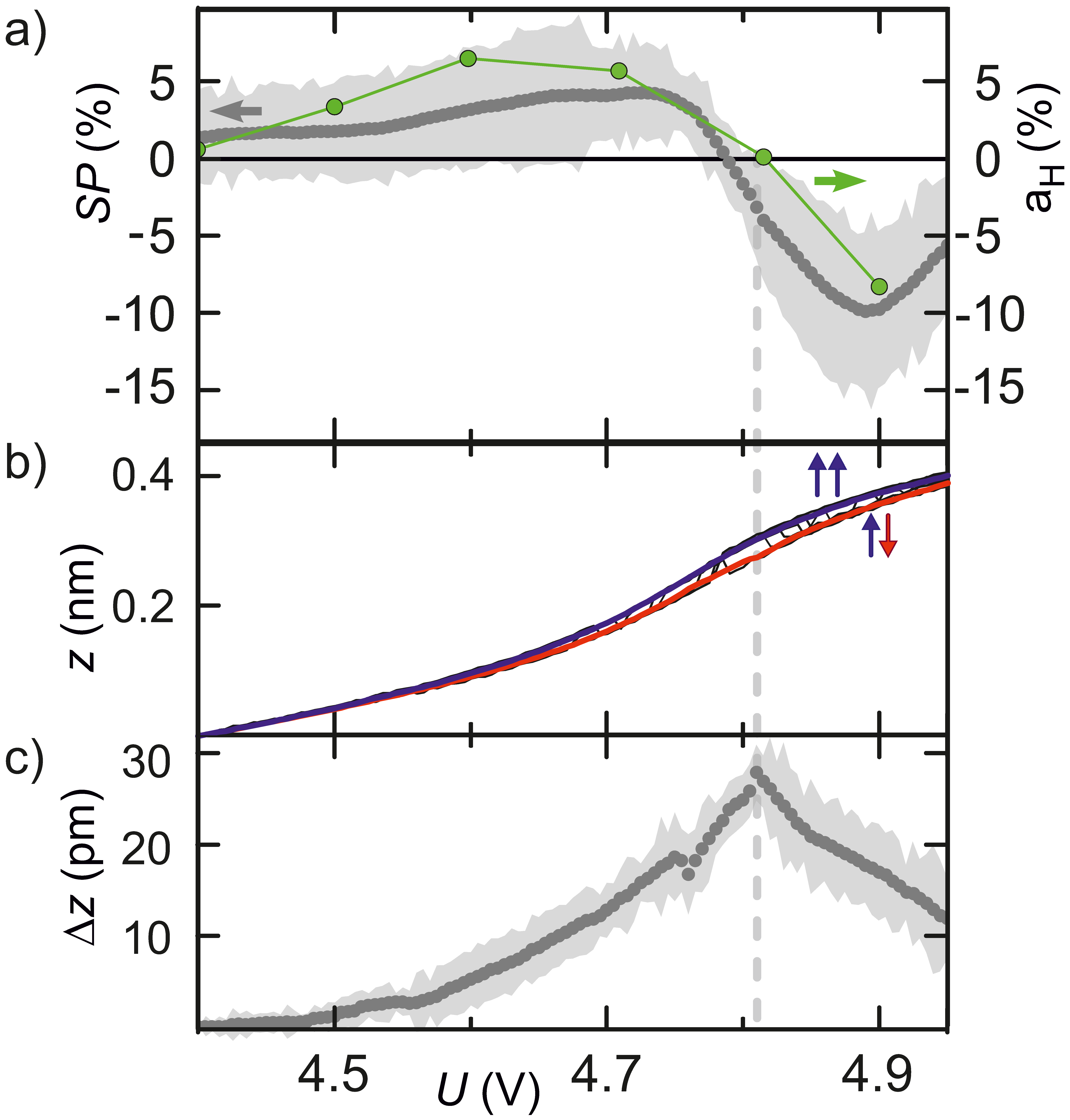}
\caption{\textbf{Spin-polarized constant current resonant tunneling spectroscopy: comparison of bias-dependent spin-resolved signals.}
\textbf{(a)}~Spin-polarization $SP(U)$ of the hot-electron gas (dark grey) with measurement uncertainty (light grey) calculated from the spin-resolved d$I$/d$U(U)$ data, as presented in Fig.~4 of the main manuscript.
Current-induced asymmetry~$a_{\rm H}$ (green) between the two magnetization states of the nanomagnet as a function of bias~$U$, as shown in Fig.~4 of the main manuscript.
\textbf{(b)}~Spin-resolved constant current $z(U)$ curves recorded on the nanomagnet's center, simultaneously with the spin-resolved d$I$/d$U(U)$ curves, presented in Fig.~4 of the main manuscript. 
$z_{\uparrow\uparrow}(U)$ curve (blue) for parallel and $z_{\uparrow\downarrow}(U)$ (red) for antiparallel configuration of tip and sample magnetization.
The data is obtained by binning a set of curves (black) recorded on the nanomagnet that rapidly switches between its two magnetic orientations.
\textbf{(c)}~Spin-polarized $\Delta z$ computed from the difference of curves in (b). 
The maximum in $\Delta z$, marked by the dashed line, coincides with the zero crossing of $SP(U)$.
($I=2$\,nA, $T=39.5$\,K.)}
\label{fig_dz}
\end{center}
\end{figure}

In SP-STM, when the STM feedback is regulating on the overall tunneling current, the spin-polarized portion of the current leads to differences in the apparent height of magnetic nanostructures imaged with magnetic tips dependant on the cosine of the enclosed angle of the tip's and sample's magnetic moments.~\cite{wiesendanger_observation_1990, wiesendanger_spin_2009}
This translates directly onto the spin-dependent constant current resonant tunneling spectra, where the relative distance~$z$ of the tip and the sample is registered alongside the differential conductance d$I/$d$U$ when ramping the sample bias~$U$.
In this case, the $\Delta z(U)$ computed from the difference of the $z(U)$ curves for parallel ($\uparrow\uparrow$) and antiparallel ($\uparrow\downarrow$) tip and sample magnetic configuration, respectively, is proportional to the total integrated spin current.
As long as the $\Delta z$ does not change its sign the total spin current integrated from $E_{\rm Fermi}$ to an energy given by the bias~$U$ also maintains the same polarity.

Figure~\ref{fig_dz}(a) shows the bias-dependent effective spin-polarization of the hot-electron gas, $SP(U)$, and the histogram asymmetry~$a_{\rm H}(U)$ of the nanomagnet's states, as defined in the main text and presented therein in Fig.~4. 
Figure~\ref{fig_dz}(b) shows the spin-dependent $z_{\uparrow\uparrow}(U)$ and $z_{\uparrow\downarrow}(U)$ curves, that have been recorded simultaneously with the respective d$I/$d$U_{\uparrow\uparrow}(U)$ and d$I/$d$U_{\uparrow\downarrow}(U)$ curves presented in Fig.~4 in the main text. 

The computed $\Delta z(U)$ is shown in Fig.~\ref{fig_dz}(c).
As discussed in the main text, $a_{\rm H}(U)$ which reveals the sign and magnitude of the spin-transfer torque (STT) surprisingly follows the evolution of $SP(U)$ rather than the $\Delta z(U)$ representing the integrated spin current. 
At $U_{\rm cross}$ close to $U=4.8$\,V, $SP(U)$ changes its sign which coincides with the maximum in the $\Delta Z$.
This coincidence is expected when integrating the $SP(U)$ to acquire the total spin current polarization. 
For $U<U_{\rm cross}$ the positive $SP(U)$ continuously adds up to an increasing total current spin-polarisation, represented by an increasing $\Delta z(U)$. 
However, for $U>U_{\rm cross}$ the negative $SP(U)$ adding to the acquired total spin current diminishes its spin-polarization, resulting in an decrease in $\Delta z(U)$. 
Notably, the sign reversal in $SP(U)$ does not result in a sign reversal of $\Delta z(U)$, but rather in its decrease.
While, the latter remains positive, $a_{\rm H}(U)$ becomes negative confirming its reliance on the $SP(U)$ rather than on the total spin current. 
In summary, the sign and magnitude of the STT exerted on the magnetic nanostructure by tunneling over the resonance state, indicated by the histogram asymmetry, coincides with $SP(U)$ rather than with the total spin current represented by~$\Delta Z$.

\bibliographystyle{naturemag}
\bibliography{STTviaIPS_supp}